\documentclass[conference]{IEEEtran}
\usepackage{verbatim}
\usepackage{graphicx,epsfig}
\usepackage{amsfonts}
\usepackage{subfigure}
\usepackage[normalem]{ulem}
\usepackage{amsmath}

\def\bi{\begin{itemize}}
\def\ei{\end{itemize}}
\def\bequ{\begin{equation}}
\def\eequ{\end{equation}}

\def\benum{\begin{enumerate}}
\def\eenum{\end{enumerate}}

\begin{document}

\title{Joint Network Coding for Interfering Wireless Multicast Networks}
\author{Jalaluddin Qureshi, Chuan Heng Foh and Jianfei Cai  \\
School of Computer Engineering\\
Nanyang Technological University, Singapore\\
\ jala0001@e.ntu.edu.sg}

\maketitle

\begin{abstract}
Interference in wireless networks is one of the key-capacity
limiting factor. The multicast capacity of an ad-hoc wireless
network decreases with an increasing number of transmitting and/or
receiving nodes within a fixed area. Digital Network Coding (DNC)
has been shown to improve the multicast capacity of non-interfering
wireless network. However recently proposed Physical-layer Network
Coding (PNC) and Analog Network Coding (ANC) has shown that it is
possible to decode an unknown packet from the collision of two
packet, when one of the colliding packet is known a priori. Taking
advantage of such collision decoding scheme, in this paper we
propose a Joint Network Coding based Cooperative Retransmission
(JNC-CR) scheme, where we show that ANC along with DNC can offer a
much higher retransmission gain than that attainable through either
ANC, DNC or Automatic Repeat reQuest (ARQ) based retransmission.
This scheme can be applied for two wireless multicast groups
interfering with each other. Because of the broadcast nature of the
wireless transmission, receivers of different multicast group can
opportunistically listen and cache packets from the interfering
transmitter. These cached packets, along with the packets the
receiver receives from its transmitter can then be used for decoding
the JNC packet. We validate the higher retransmission gain
performance of JNC with an optimal DNC scheme using simulation.

\end{abstract}

\section{Introduction} \label{sect:Introduction}
Wireless multicasting is seen as a bandwidth efficient mean of
disseminating common information to multiple receivers. Various
emerging applications such as wireless multi-player
gaming~\cite{Kondo09} and multimedia broadcast (currently being
standardised by the IEEE 802.11aa working group) are based on
wireless multicasting. While a previous empirical
study~\cite{Akella05} on the deployment of wireless Access Points
(AP) in metropolitan areas has shown that APs are often deployed in
a chaotic manner, with several of these APs therefore often
competing with each other for access to the same transmission
channel. Therefore with an increase in wireless mutlicast data
traffic, and an increasing density of wireless nodes within a fixed
area competing for the same channel, an efficient solution is needed
which addresses both the increasing wireless multicast bandwidth
demand and the constraint of wireless interference.

Digital network coding (DNC) has been shown to be one such technique
which improve the capacity of multicast wireless
network~\cite{Karande11}, and its reliability
gain~\cite{Nguyen09}~\cite{Qureshi09} for a non-interfering network.
In DNC, multiple packets are coded together over Galois Field
$GF(q)$, where $q$ is the field size. If the coding vector is
randomly selected from the Galois Field, then such DNC scheme is
known as Random Linear Network Coding (RLNC)~\cite{Kondo09} scheme.
If this coding vector is selected deterministically then such DNC
scheme is known as deterministic network coding, the most commonly
used form of the deterministic network coding is known as
XOR-coding~\cite{Nguyen09}~\cite{Qureshi09}, i.e. deterministic
network coding over $GF(2)$.

Recent works have shown thats network coding can also be performed
at the physical layer. Such network coding performed at physical
layer, known as Physical-layer Network Coding (PNC)~\cite{Zhang06}
and Analog Network Coding (ANC)~\cite{Katti07} can improve the
throughput order for multi-pair unicast transmission in ad-hoc
wireless network~\cite{Chen08}. In PNC/ANC, a node can decode the
unknown packet $c_2$ from the collided packet $c_1\odot
c_2$\footnote{We use the $\odot$ notation to denote the collision
operation, and $\oplus$ to denote XOR-coding operation.}, provided
that the node has packet $c_1$ a priori. The key difference between
PNC and ANC is that, PNC requires the colliding packets to collide
in a perfectly synchronised manner, whereas in ANC, the colliding
packet need not necessarily collide perfectly synchronised, which
therefore makes ANC a more practical coding scheme for
implementation. Therefore in our implementation we use the ANC
scheme.

While a significant amount of work has been done to characterise the
throughput benefits of DNC~\cite{Karande11}~\cite{Nguyen09} and
ANC~\cite{Chen08}~\cite{Qureshi11} in isolation, to the best of our
knowledge there has been no work done so far which characterises the
joint benefits of DNC and ANC. Further, so far, ANC applications has
only been limited for simple relay networks. Therefore our current
work is also the first work of its kind extending the application of
ANC beyond relay networks.

In this paper we demonstrate the throughput gain by jointly using
ANC and DNC (deterministic XOR-coding) to transmit packets to
receivers in a single-hop setting where two wireless multicast group
interfere with each other, which we call Joint Network Coding using
Cooperative Retransmission (JNC-CR). The rest of the paper is
organsied as follow. We present an overview of related works in
Section~\ref{sect:Related}. In Section~\ref{sect:Model}, we
characterise the system model of the network. We then present the
JNC-CR protocol design, along with an illustrating example in
Section~\ref{sect:Protocol}, followed by simulation results in
Section~\ref{sect:Simulation} and summary of our work in
Section~\ref{sect:Conclusion}.

\section{Related Work} \label{sect:Related}
\begin{figure}
\begin{center}
\includegraphics[width = 0.5\textwidth]{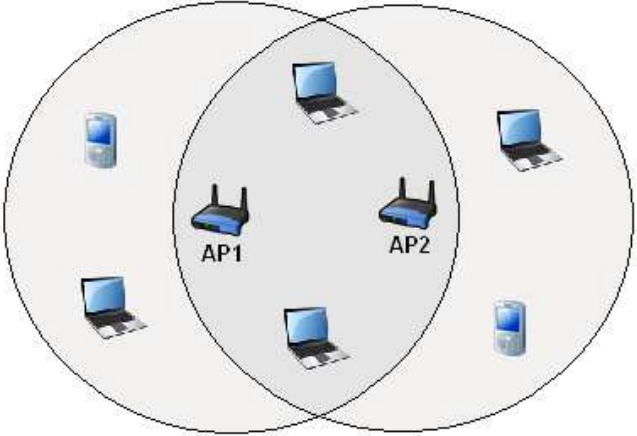}
\end{center}
\caption{Two interfering multicast network, with N=3 and M=1.}
\label{fg:fig1}
\end{figure}

Our current work is primarily an extension of our previous work,
Cooperative Retransmission (CR) through collission~\cite{Qureshi11}.
While in our previous work we had shown the retransmission gains of
utilising ANC over IEEE 802.11 ARQ (Automatic Repeat Request) for 2
unicast transmissions interfering with each other, in our current
work, we demonstrate the retransmission gains of JNC-CR over an
optimal DNC scheme for 2 multicast transmissions interfering with
each other.

\subsection{DNC Retransmission scheme}
A bulk of previous
works~\cite{Kondo09}~\cite{Nguyen09}~\cite{Qureshi09}~\cite{Rozner07}~\cite{Rouayheb07}
on network coding based retransmission schemes have been limited
towards using DNC, comparing the performance gain of DNC over ARQ
retransmission schemes. Advantages in such coding gains come from
the independent Bernoulli packet loss model in wireless networks, as
experimentally shown~\cite{Salyers08}. In~\cite{Salyers08} the
authors showed that while the average packet loss probability of 2
co-located identical receivers are similar, the receivers' packet
loss burstiness may however differ under similar conditions. Hence,
instantaneous packet loss for 2 such receivers receiving
transmission from the same transmitter show no correlation pattern.
Therefore for a transmitter $AP_1$ multicasting packet $c_1$ and
$c_2$ to receivers $R_1$ and $R_2$, $c_1$ may be received by $R_1$
only while $c_2$ may be received by $R_2$ only, however unlike ARQ
retransmission scheme where both the packets are retransmitted
separately in 2 different time slots, in a DNC based retransmission
scheme the packets $c_1$ and $c_2$ are XOR-coded as $c_1\oplus c_2$
and retransmitted in 1 time slot. Upon reception of this coded
packet both $R_1$ and $R_2$ can decode the coded packet using the
packet they had received earlier. Therefore DNC reduces the time
slots needed to retransmit the lost packets from 2 time slots to 1
time slots for this example. Our previous work, CR~\cite{Qureshi11}
was the first work of its kind to improve retransmission gain using
ANC rather than DNC. CR is implemented by two interfering APs using
the common superimposed acknowledgement information, transmitted by
the receivers in the interference region.

\subsection{Superimposed Acknowledgement}
Superimposed acknowledgement~\cite{Foh10}~\cite{Durvey07} is an
ACK-thinning scheme for multicast transmission. In a superimposed
acknowledgement each receiver in the multicast network transmits a
unique ACK packet, which is embedded with a unique predefined
bitstream patterns, such that the simultaneous collision of these
packets will result in a different collided packet for different
permutation of ACK packets colliding together. Therefore all
receivers which have received the packet transmit their ACK packet
in the same time slot. This collided packet can then be used to
determine which set of receivers have received the data packet.
Because of the broadcast nature of these collided ACK packet by
receivers in the interference region, both the APs have the packet
reception status of these receivers.

\subsection{Cooperative Retransmission}
In a CR~\cite{Qureshi11} scheme, 2 interfering APs retransmit
selected lost packets simultaneously, resulting in a collided
packet, which is then decoded by the receivers using ANC and packets
opportunistically overheard from the interfering AP. Such
cooperative retransmission scheme is implemented without any complex
handshaking or scheduling procedure. Consider, for example unicast
transmissions $AP_1 \sim R_1$ and $AP_2 \sim R_2$, interferring with
each other and $R_1$ and $R_2$ located such that both the receivers
can hear transmission from $AP_1$ and $AP_2$. Because of the
broadcast nature of wireless transmission, it is therefore possible
that $R_1$ may overhear (also known as opportunistic listening)
packet destined for $R_2$ and vice-versa. We use superimposed
acknowledgement in such a CR scheme, wherein both $R_1$ and $R_2$
simultaneously transmit ACK packet for transmission received by both
$AP_1$ and $AP_2$. Therefore both the APs are aware of packet
reception status of $R_1$ and $R_2$. In such a network packet $c_1$
transmitted by $AP_1$ destined for $R_1$ can be overheard by $R_2$
but not by $R_1$, whereas packet $c_2$ transmitted by $AP_2$ for
$R_2$ can be overheard by $R_1$ but not by $R_2$. Because of the
superimposed acknowledgement scheme, $R_2$ will transmit ACK packet
for $c_1$, and $R_1$ will transmit ACK packet for $c_2$. $AP_1$ and
$AP_2$ can then simultaneously retransmit $c_1$ and $c_2$ in the
same time slot which results in the collided packet $c_1\odot c_2$.
Each of the receiver can then use the previously, opportunistically
received packet to decode $c_1\odot c_2$ using ANC. This therefore
reduces the total number of retransmissions needed from 2 to 1 for
this example. In this work we further expand on the benefits of both
ANC and DNC, and show that using JNC-CR offers a much higher
retransmission gain.

\section{System Model} \label{sect:Model}
Consider two APs, $AP_1$ and $AP_2$ multicasting packets to
receivers in their network. Let $d_{AP}$ denote the distance between
the two APs, and $r_t$ denote the transmission range of each AP,
characterized as an omnidirectional radio propagation, with both the
APs following the uniform power assignment scheme, i.e. transmitting
at an equal transmission power, and at the same transmission rate.
Consider that the interfering APs are overlapped such that $d_{AP}<
2r_t$. Each AP associates with $N$ client stations, which are
uniformly distributed within the transmission range of the AP.
Average packet loss probability $p_{ij}$ for transmissions from
$AP_i$ to receiver $R_j$ follows an independent Bernoulli packet
loss model~\cite{Salyers08}, where $1\leq i \leq 2$ and $1\leq j
\leq 2N$. Receivers $1\leq j \leq N$ are connected to $AP_1$ and
receivers $N< j \leq2N$ are connected to $AP_2$. The total number of
nodes in the overlap region is given as $2M$, where $1\leq M \leq
N$. We assume uniform packet loss $p$ for both the network, such
that $p_{ij}=p, \forall i, j$. The probability that a receiver in
the non-interference region correctly receives a packet is given as
$(1-p)$, whereas the probability that a receiver in the interference
region correctly receives a collided packet is given as
$(1-p)^2$~\cite{Qureshi11}, as both the colliding packet need to
reach correctly at the receiver. Packet batch size for transmissions
by $AP_i$ is denoted as $B_i$, for simplicity we assume that
$B_1$=$B_2$=$B$. For multimedia applications such as video streaming
and file sharing, $B$ is usually a large value. We assume a reliable
feedback mechanism, this is consistent with the previous assumptions
used in similar
works~\cite{Nguyen09}~\cite{Qureshi09}~\cite{Rozner07}. Further we
also assume a reliable superimposed acknowledgement feedback
mechanism in our protocol design, consistent with our previous
work~\cite{Qureshi11}.

\subsection{Performance overview}
In our performance evaluation, we compare JNC-CR with an optimal DNC
scheme. However since it is a NP-hard problem to decide whether an
optimal number of transmission can be achieved for a given field
$GF(q)$~\cite{Rouayheb07}, we therefore assume an infinitely large
value of $q$ for DNC. Without loss of ambiguity we ignore the large
coding overhead of $Blog_2(q)$ bits for DNC corresponding to an
infinitely large value of $q$. The expected number of transmissions
(and hence retransmissions) needed to transmit $B$ packets to $N$
receivers using DNC has been calculated in Equation 13
of~\cite{Sagduyu07}. We therefore use result from~\cite{Sagduyu07},
with $q=\infty$ as a lower bound for DNC, for specified values of
$p$, $B$ and $N$.

Packet overhead of a XOR coded packet is given as $Nlog_2(q)$, where
$q=2$. Implementing ANC has a total packet overhead of 128
bits~\cite{Katti07}. Therefore the total packet overhead of a JNC
packet is given as $(128+2N)$ bits.

\section{JNC CR Protocol Design} \label{sect:Protocol}
In JNC, packets are encoded at two level. In the first instance,
each $AP$ XOR selected data packets bit-by-bit, and then both the
APs simultaneously transmit the XOR coded packet, which results in
ANC of the XOR coded packets. When a receiver receives a JNC packet,
it first performs ANC decoding on the JNC packet to retrieve the
XOR-coded packets. Once a receiver decodes the collided packet it
then retrieves the XOR-coded packet, on which it then performs XOR
decoding to retrieve the data packets.

The APs make XOR coding decision using BENEFIT coding
algorithm~\cite{Qureshi09}, and ANC coding decision using a simple
ANC-CR coding decision as shown in Table~\ref{table:pseudocode}.
BENEFIT is a memory based heuristic coding algorithm which makes
coding decision such that every receiver receives an innovate packet
on the reception of the coded packet. An innovative packet, is a
packet which the receiver can not generate using the set of packets
it already has.

\subsection{Illustrating example} \label{sect:Example}

\begin{table}[htbp]
\caption{Transmission Matrix Example} \label{table:matrix}
\begin{center}
\begin{tabular}{|c|c|c|c|c|}
\hline             & $c_1$ & $c_2$ & $c_3$ & $c_4$  \\
\hline $R_1$       &   1   &   0   &   0   &   0    \\
\hline $R_2$       &   0   &   1   &   -   &   -    \\
\hline $R_3$       &   0   &   0   &   0   &   1    \\
\hline $R_4$       &   -   &   -   &   1   &   0    \\
\hline
  \end{tabular}
\end{center}
\end{table}

Consider for illustration a simple example where $AP_1$ is
multicasting packets $c_1$ and $c_2$ to $R_1$ and $R_2$, while
$AP_2$ is multicasting packets $c_3$ and $c_4$ to $R_3$ and $R_4$.
$R_1$ and $R_3$ are located in the interference region, whereas
$R_2$ and $R_4$ are located in the non-interference region. The
reception status of each packet is given in
Table~\ref{table:matrix}, where `1' represents that the packet has
not been received by the corresponding receiver, `0' represents that
the packet has been received, while `-' denotes that the receiver is
not within the transmission range of the AP transmitting that
packet. In an ARQ based retransmission scheme, $AP_1$ and $AP_2$
will retransmit these lost packets in different time slots,
therefore requiring a total of 4 time slots to retransmit all the 4
packets. In a DNC based scheme, $AP_1$ and $AP_2$ transmit the
encoded packet $c_1\oplus c_2$ and $c_3\oplus c_4$ respectively
which the receivers can decode using the packet each receiver
already has. Therefore a DNC based retransmission scheme requires a
total of 2 time slots. JNC further improves on the retransmission
gain by allowing both the APs to simultaneously retransmit
$c_1\oplus c_2$ and $c_3\oplus c_4$, which results in the 2-layer
encoded packet $(c_1\oplus c_2) \odot (c_3\oplus c_4)$ received at
$R_1$ and $R_3$, while receiver $R_2$ and $R_4$ receive the
XOR-coded packet $c_1\oplus c_2$ and $c_3\oplus c_4$ respectively,
as these receivers do not fall in the interference region. Since
$R_1$ has packet $c_3$ and $c_4$, it can use these packets to
generate $c_3\oplus c_4$ and decode $(c_1\oplus c_2) \odot
(c_3\oplus c_4)$ using ANC decoding, it then performs XOR decoding
to retrieve packet $c_1$ from $c_1\oplus c_2$. Therefore in a JNC
based retransmission scheme, a total of 1 time slots are needed to
retransmit the lost packet. Therefore for this simple example JNC
provides a retransmission gain of 4 over ARQ, and 2 over DNC.

\subsection{Non-Cooperative Collision Coding}
Each AP is only aware of the packet reception status of the
receivers located within its transmission range. In our model, both
the APs start the retransmission phase after transmitting $B$
packets using IEEE 802.11 based Carrier Sense Multiple Access
Collision Avoidance (CSMA/CA). The retransmission process take place
in 2 stages, the first stage is non-cooperative packet transmission,
whereas the second stage is cooperative packet transmission. In the
first stage, since the interfering AP is not aware of the packet
reception status of receivers not within its transmission range,
both the AP make independent DNC coding decisions. Receivers in the
non-interference region receive an XOR coded packet, whereas
receivers in the interference region receive a collided XOR coded
packet.

However since each of the AP make such coding algorithm decisions
independently, receivers in the interference region may not
necessarily benefit from such transmissions. This is because,
receivers in the non-interference region only need to perform XOR
decoding, whereas receivers in the interference region need to
perform both ANC and XOR decoding. So while each AP can make coding
decision such that the coded packet can be XOR decoded by every
receiver in the multicast group of that AP, receivers in the
interference region may not necessarily be able to perform ANC
decoding of the collided packet from the interfering AP.

In the non-cooperative collision retransmission phase receivers in
the interference region can therefore perform collision decoding
\emph{opportunistically}. Let $k$ represent the cardinality of the
XOR-coded packet, BENEFIT coding algorithm~\cite{Qureshi09} is
designed such that $1\leq k\leq N$. The probability that a receiver
can opportunistically perform collision decoding is given as the
product of the probability it receives a correct collided packet,
and the probability that it has already opportunistically overheard
the $k$ packets from the interfering AP previously, $(1-p)^{2+k}$.
Therefore given the higher packet reception probability for
receivers in the non-interference region, receivers in the
non-interference region recover the lost packets much earlier than
the receivers in the interference region. Once all the receivers in
the non-interference region have correctly received the lost
packets, the APs can then perform Cooperative Collision Coding.

\subsection{Cooperative Collision Coding}
\begin{table}
\caption{ANC-CR coding algorithm, Pseudocode}
\label{table:pseudocode}
\begin{tabular}{|l|}
\hline

$c_{ni}\longleftarrow$ Coded packet generated by $AP_i$\\
$n_i\longleftarrow$ Number of receivers for which $c_{ni}$ is an innovative packet\\

\\
\textbf{for} (m=1; m$\leq$2M; m++)\\
\ \ \ \ \textbf{if} (node m can perform collision decoding of
$c_{n1}\odot c_{n2}$)\\
\ \ \ \ \ \ collision decoding++\\
\\
JNC benefit = collision decoding $\cdot$ $(1-p)^2$\\
$AP_i$ benefit = $n_i \cdot (1-p)$, $\forall i$\\
XOR benefit = max($AP_1$ benefit, $AP_2$ benefit)\\
\\
\textbf{if} (JNC benefit$>$XOR benefit)\\
\ \ \ \ Simultaneously transmit coded packet, collision-coding\\
\textbf{else}\\
\ \ \ \ AP with higher XOR benefit transmits XOR-coded packet,\\
\ \ \ \ collision-free\\

\hline
\end{tabular}
\end{table}

In the Cooperative Collision phase, only the receivers in the
interference region need to recover the lost packets. Because of the
broadcast nature of superimposed acknowledgement, both the APs are
aware of all the packet reception status of all the receivers in the
interference region, and since both the APs run the same ANC-CR
coding algorithm, both the APs are also aware of the coding decision
the interfering AP makes. A pseudocode of the ANC-CR algorithm is
given in Table~\ref{table:pseudocode}. Both the APs simultaneously
run the same coding algorithm, and weigh in the benefit of
simultaneously transmitting the coded packet. If the benefit of
simultaneously transmitting the coded packet is greater than the
benefit of transmitting either of the coded packet without
collision, then both the APs transmit their coded packet
simultaneously, which result in the collision of the coded packet.
If however the benefit of transmitting an XOR from either AP is
higher than JNC-CR, then the AP with higher transmission benefit
transmit the packet. As we had shown in~\cite{Qureshi11} such
cooperative retransmission can be implemented without any complex
handshaking or scheduling procedures.

\section{Simultation} \label{sect:Simulation}
\begin{figure}
\begin{center}
\includegraphics[width = 0.5\textwidth]{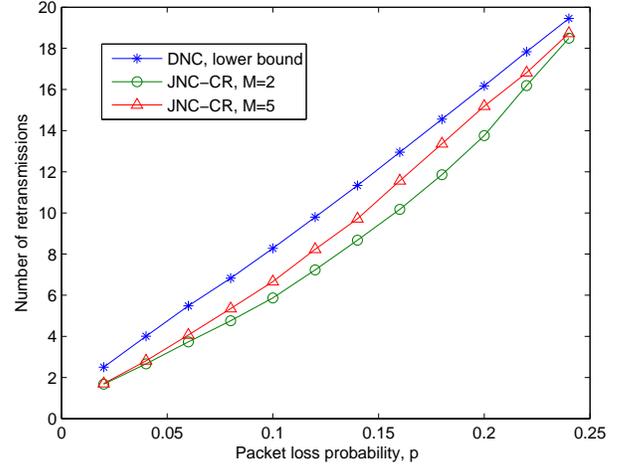}
\end{center}
\caption{Average number of retransmission under different $p$
values, for N=5 and B=20.} \label{fg:fig1}
\end{figure}
\begin{figure}
\begin{center}
\includegraphics[width = 0.5\textwidth]{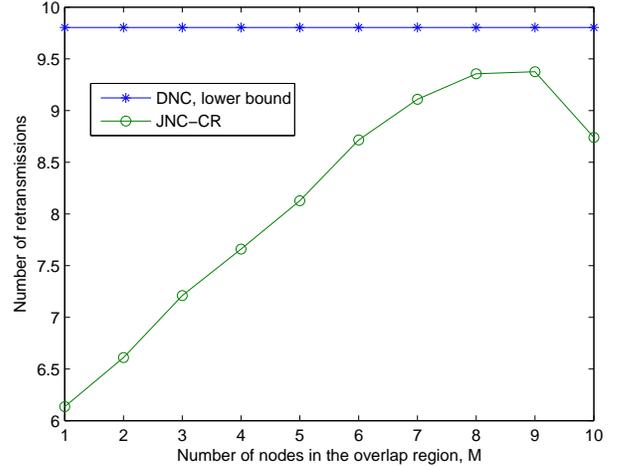}
\end{center}
\caption{Average number of retransmission under different $M$
values, for N=10, p=0.1 and B=20.} \label{fg:fig2}
\end{figure}

Packet decoding using ANC has been successfully demonstrated on a
test bed in~\cite{Katti07}. Therefore we can assume that ANC is a
practically applicable technique. For the proposed collision based
cooperative retransmission, we construct a C++ discrete-time
simulator with the system model described in
Section~\ref{sect:Model}.

Fig.~\ref{fg:fig1} shows the average number of retransmission for
different $p$. As $p$ increases the number of retransmission also
increases for both DNC and JNC-CR, consistent
with~\cite{Nguyen09}~\cite{Rozner07}~\cite{Sagduyu07}. JNC-CR shows
retransmission gain over DNC. The higher gain for decreasing $M$ in
Fig.~\ref{fg:fig1} and~\ref{fg:fig2} comes from the cooperative
collision coding stage. The probability that the receiver in the
interference region will be able to perform ANC-decoding is given as
$(1-p)^{2+k}$, for cooperative collision coding, $1\leq k\leq M$.
Therefore for $M=2$ more packets get ANC-coded compared to $M=5$,
which improves the retransmission gain. For Fig.~\ref{fg:fig2}, a
sudden dip in the number of retransmission occurs for $M=10$,
because all the packets are then retransmitted in a cooperative
collision coding, and results in a more efficient ANC coding
decision.

\begin{figure}
\begin{center}
\includegraphics[width = 0.5\textwidth]{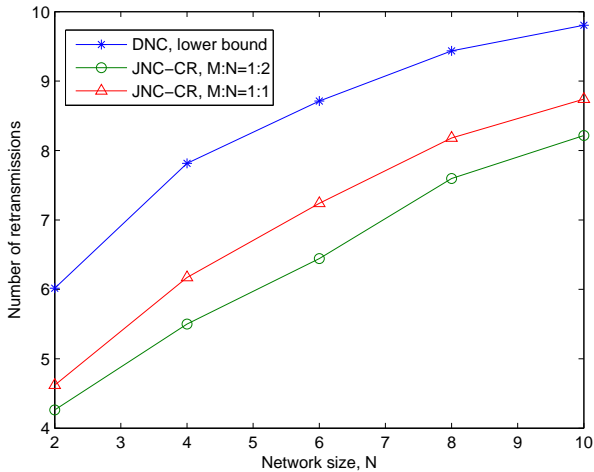}
\end{center}
\caption{Average number of retransmission under different $N$
values, for p=0.1 and B=20.} \label{fg:fig3}
\end{figure}

\begin{figure}
\begin{center}
\includegraphics[width = 0.5\textwidth]{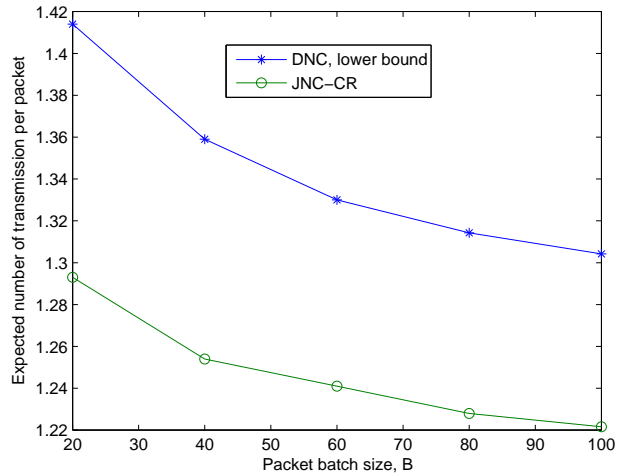}
\end{center}
\caption{Expected number of transmissions per packet for different
packet batch size, for p=0.1, N=5 and M=2.} \label{fg:fig4}
\end{figure}

Fig.~\ref{fg:fig3} shows that the average number of retransmission
increases logarithmically as the network size increases.
Fig.~\ref{fg:fig4} shows that the average number of retransmissions
decreases logaritmically for increasing packet batch size. However
using a large packet batch size will increase transmission latency.
The results of Fig.~\ref{fg:fig3} and ~\ref{fg:fig4} are consistent
with~\cite{Nguyen09}~\cite{Rozner07}~\cite{Sagduyu07}. For both
these figures, JNC-CR shows better retransmission bandwidth than
DNC.

\section{Conclusion} \label{sect:Conclusion}
In this work we have demonstrated the retransmission bandwidth gain
of JNC-CR over DNC. Such a scheme can be implemented without any
complex handshaking or scheduling procedure, and address the current
wireless bandwidth demand, and increasing density of wireless
networks in metropolitan areas. JNC-CR can also address the
intra-flow interference problem in wireless multicast routing.

\bibliographystyle{IEEEtran}
\bibliography{IEEEabrv,icics}

\end{document}